\documentclass[conference]{IEEEtran}
\IEEEoverridecommandlockouts

\usepackage{cite}
\usepackage{amsmath,amssymb,amsfonts}
\usepackage{algorithmic}
\usepackage{graphicx}
\usepackage{textcomp}
\usepackage{xcolor}
\usepackage{booktabs}
\usepackage{url}
\usepackage{tikz}
\usetikzlibrary{arrows.meta,positioning,fit,backgrounds,shapes.geometric,calc}
\graphicspath{{assets/}{./assets/}}

\newcommand{\repourl}{https://github.com/agarnung/FrAM-Motion-Magnification}
\def\BibTeX{{\rm B\kern-.05em{\sc i\kern-.025em b}\kern-.08em
    T\kern-.1667em\lower.7ex\hbox{E}\kern-.125emX}}

\begin{document}

\title{Fractional-Order Adaptive Motion Magnification:\\
Phase-Reliability Weighting for Noise-Constrained\\
Video Amplification}

\author{
    \IEEEauthorblockN{Alejandro Garnung Menéndez, M.Sc., PhD Candidate\textsuperscript{1}}
    \IEEEauthorblockA{
        \textsuperscript{1}\textit{University of Oviedo, Gijón Polytechnic School of Engineering, Asturias, Spain} \\
        \texttt{uo269564@uniovi.es}, \texttt{garnungalejandro@gmail.com}
    }
}

\maketitle

\begingroup
\renewcommand{\thefootnote}{}
\footnotetext{Code, the interactive real-time viewer and result videos are
available at \expandafter\url\expandafter{\repourl}.}
\addtocounter{footnote}{-1}
\endgroup

\begin{abstract}
Eulerian video amplification boosts sub-pixel motion by band-pass filtering
per-pixel intensity traces and applying a uniform gain. That gain ignores local
structure, so sensor noise is amplified together with the signal, especially in
textureless regions where the monogenic phase is unreliable. We propose FrAM
(Fractional-order Adaptive Motion Magnification), a pipeline developed first
offline and then as a causal stream. It replaces the constant temporal gain
with a Gr\"unwald--Letnikov derivative of fractional order, giving continuous
control over high-frequency emphasis, and replaces the uniform spatial gain with
a per-pixel weight derived from the local amplitude of the monogenic signal.
On a controlled synthetic sequence split into textured and flat halves, FrAM
matches the amplification of the Eulerian baseline while keeping flat-region
temporal noise at the input level. The reduction holds across an eightfold range
of input noise levels. Real videos show improved spatial selectivity and lower
background noise in every case. The causal reformulation cuts the per-frame cost
by two orders of magnitude, reaching 69\,fps at 640$\times$480.
\end{abstract}


\section{Introduction}

A camera records many motions that are too small for a human observer to see:
the pulse-induced undulation of skin, the vibration of a mechanical structure,
or the sway of a loaded beam. The purpose of \emph{video magnification} is to
make these imperceptible displacements visible without attaching physical
sensors to the scene.

The Eulerian formulation introduced by Wu \emph{et al.}~\cite{wu2012} recasts
the problem in a way that avoids the difficulties of tracking. Instead of
following points across frames---a Lagrangian strategy that requires
sub-pixel correspondence and is fragile at fine scales---it monitors the
intensity evolution at \emph{fixed} pixel locations. For a small displacement
$\delta(t)$, a first-order Taylor expansion gives
\begin{equation}
I(x + \delta(t), t) \approx I(x,t) + \delta(t)\,\nabla I(x,t),
\label{eq:taylor}
\end{equation}
so motion manifests as an intensity fluctuation proportional to the local
spatial gradient. What was a spatial estimation problem becomes a temporal
filtering problem: decompose each frame into spatial-frequency bands with a
Laplacian pyramid, temporally band-pass each band to isolate the frequency of
interest, multiply by a gain $\alpha$, and reconstruct.

The limitation of this approach is that \eqref{eq:taylor} gives no indication
of \emph{where} the amplification should be applied. The gain is the same for
every pixel, so the temporal band-pass amplifies sensor noise and genuine
motion with identical strength. Because noise spans the whole spectrum, some of
it always lies inside the pass-band. In regions that lack structure the output
is therefore mostly amplified noise, producing the temporal mottling commonly
reported in the literature. In practice it is this artefact, not any formal
upper bound, that restricts the useful gain.

Phase-based methods~\cite{wadhwa2013,wadhwa2014} improve the situation by
amplifying the \emph{phase} of a complex steerable or Riesz pyramid rather
than its amplitude, shifting structure without changing contrast. Their noise
behaviour is markedly better, but an important precondition is often left
unstated: the local phase of the monogenic signal~\cite{felsberg2001} is well
defined only where the local amplitude is non-negligible. Where
$A \to 0$ the phase collapses into noise, and amplifying it reintroduces
the very artefact the phase formulation was designed to avoid.

This paper builds on that observation and makes two contributions:

\begin{enumerate}
\item A \textbf{fractional-order temporal filter} based on the
Gr\"unwald--Letnikov derivative, applied after band-pass filtering with the
frequency response $|(1-e^{-i\omega})^{\nu}|$. The order
$\nu \in [0,1]$ continuously interpolates between the identity ($\nu=0$) and a
first temporal difference ($\nu=1$), setting how strongly high temporal
frequencies---and therefore noise---are emphasised.

\item A \textbf{phase-reliability adaptive gain} that modulates the
amplification per pixel according to the local monogenic amplitude. Where the
phase is uninformative the gain is reduced. The cutoff $\sigma$ is chosen as a
percentile of the amplitude distribution of each band, so the method adapts to
the content without manual tuning.
\end{enumerate}

The evaluation uses a synthetic sequence whose ground-truth displacement is
known, which makes it possible to measure the achieved amplification factor
rather than relying solely on visual inspection. An ablation is included to
disentangle the two contributions.

\section{Related Work}

\subsection{Amplitude-based magnification}
Eulerian Video Magnification~\cite{wu2012} established the pipeline just
described. Its simplicity and low computational cost have made it the baseline
method, and it remains effective for colour magnification, where the quantity
of interest is an intensity change rather than a displacement. Its main
weakness is the linear growth of noise with the gain.

\subsection{Phase-based magnification}
Wadhwa \emph{et al.}~\cite{wadhwa2013} replaced the Laplacian pyramid with a
complex steerable pyramid~\cite{simoncelli1995} and amplified local phase,
achieving much better noise performance at higher computational cost. The Riesz
pyramid~\cite{wadhwa2014} lowered that cost by approximating the quadrature
pair with the Riesz transform, a two-dimensional extension of the Hilbert
transform, and reached real-time speeds. The multipliers that define the Riesz
transform belong to the family of singular integrals studied in fractional
calculus~\cite{podlubny1999}, a link that motivated the present formulation.

\subsection{Fractional calculus in image processing}
Fractional-order operators have been used for image denoising, edge detection,
and texture enhancement~\cite{pu2010,yang2016}. In those applications the
non-integer order provides a trade-off between edge sharpening and noise
suppression that integer-order derivatives cannot offer. Almost all such work
is \emph{spatial}. Applying fractional differentiation along the
\emph{temporal} axis of a video-magnification pipeline, as we do here, does
not appear to have been explored before.

\section{Method}

\subsection{The Gr\"unwald--Letnikov operator}

Among the many equivalent definitions of a fractional
derivative~\cite{podlubny1999}, the Gr\"unwald--Letnikov form is the most
convenient to discretise directly:
\begin{equation}
D^{\nu} f(t) \approx h^{-\nu} \sum_{k=0}^{K} w_k \, f(t - kh),
\qquad w_k = (-1)^k \binom{\nu}{k},
\label{eq:gl}
\end{equation}
with coefficients computed through the numerically stable recurrence
\begin{equation}
w_0 = 1, \qquad w_k = w_{k-1}\,\frac{k-1-\nu}{k}.
\label{eq:recurrence}
\end{equation}

Two properties of \eqref{eq:gl} are central to the method.

\begin{figure}[t]
\centering
\includegraphics[width=\columnwidth]{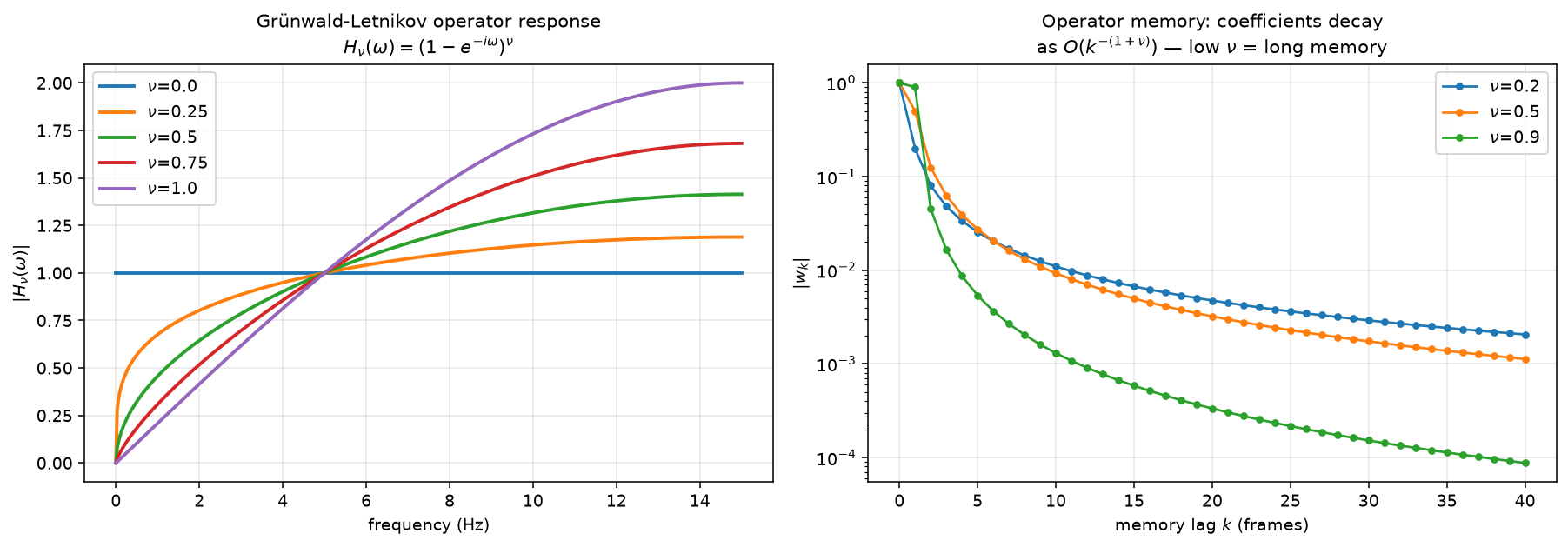}
\caption{Left: magnitude of the Gr\"unwald--Letnikov response
\eqref{eq:response}, showing the continuous interpolation between the identity
($\nu=0$) and a first derivative ($\nu=1$). Right: decay of the coefficients
$w_k$ as a function of memory lag $k$; smaller $\nu$ implies a longer effective
memory. Here $k$ indexes past frames in the GL sum and measures the operator's
temporal reach, not a group-delay or control-loop reaction lag.}
\label{fig:operator}
\end{figure}

\textbf{Continuous interpolation.} For $\nu = 1$ the coefficients become
$[1,-1,0,\dots]$, which is the classical backward difference
$x[n]-x[n-1]$; for $\nu = 2$ they become $[1,-2,1]$, the classical
second-order backward difference. The frequency-domain expression follows from
the shift property of the discrete Fourier transform: a one-sample delay,
$x[n-1]$, corresponds to multiplication by $e^{-i\omega}$, so the first-order
backward difference has response
\[
H_1(\omega)=1-e^{-i\omega}.
\]
Raising this response to an arbitrary real power gives the continuous-order
generalisation
\begin{equation}
H_{\nu}(\omega) = \left(1 - e^{-i\omega}\right)^{\nu}
\sim (i\omega)^{\nu},
\label{eq:response}
\end{equation}
where the approximation uses $e^{-i\omega}\approx 1-i\omega$ for small
$\omega$. Integer values recover the usual finite differences, while
non-integer $\nu$ defines operators with no integer-order counterpart. Since
$(i\omega)^\nu = |\omega|^\nu e^{i\nu\pi/2}$ for positive frequencies, the
operator has a gain proportional to $|\omega|^\nu$ and a phase shift of
$\nu\pi/2$. Thus the order gives continuous control over the emphasis of
high-frequency components (Fig.~\ref{fig:operator}, left).

\textbf{Long memory.}
The Gr\"unwald--Letnikov weights decay algebraically,
$w_k = O(k^{-(1+\nu)})$, rather than exponentially~\cite{Scherer2011}. The
operator is therefore non-local in time, and the memory grows as $\nu$
decreases. Because the tail is algebraic there is no sharp cutoff; truncation
is a matter of tolerance rather than convergence. The implementation
estimates the needed length by choosing the smallest $K$ for which the
residual weight mass inside a $512$-term window falls below a relative
tolerance of $10^{-3}$. This yields $K=503$ at $\nu=0.2$ and only $K=124$ at
$\nu=0.9$ (Fig.~\ref{fig:operator}, right). At $\nu=0.2$ the residual only
crosses the tolerance near the edge of the window, so no much shorter memory
satisfies the criterion. This long memory is accepted by the offline
formulation, which works with the full temporal history; Sec.~VII and
Table~\ref{tab:memory} show that a causal implementation needs far less.

\subsection{Fractional band-pass filter}

Because the fractional operator emphasises high frequencies, applying it
directly to a broadband signal would also boost out-of-band noise. The
temporal processing is therefore split into two stages. A second-order
Butterworth band-pass, implemented as a biquad (two poles and two zeros;
Appendix~\ref{app:impl} explains why the order-$1$ design used here produces
such a section), restricts each band-level time series to the interval
$[f_{lo}, f_{hi}]$. It is applied with zero-phase forward--backward filtering
so that no group delay is introduced. The fractional derivative
\eqref{eq:gl} is then applied to the band-limited signal. With the sampling
interval normalised to one frame, the $h^{-\nu}$ factor in \eqref{eq:gl} is
unity and the meaning of $\alpha$ is the same for every order.

In classical Eulerian magnification~\cite{wu2012} and in phase-based
methods~\cite{wadhwa2013,wadhwa2014} the pipeline stops after the band-pass:
the result is multiplied by the constant gain $\alpha$ and added back. That
step is equivalent to applying $D^{\nu}$ with $\nu=0$, the identity. The
choice is therefore implicit and fixed, and earlier methods have no parameter
taking this role. Applying $D^{\nu}$ with $\nu \in (0,1]$ continuously
reshapes the same band-limited signal, from the original signal up to a full
temporal derivative at $\nu=1$, progressively emphasising the upper part of
the pass-band. The spectral shaping that was previously implicit becomes an
explicit, continuously tunable parameter.

\subsection{Local amplitude and phase}

For each pyramid band $\ell$, let $b$ denote the band signal of dimensions
$H_\ell \times W_\ell$. A computationally simple approximation of the two
spatial Riesz components is obtained with first-order central differences,
implemented as convolutions with the kernel $[-0.5,0,0.5]$ and its transpose.
This gives $R_1,R_2\in\mathbb{R}^{H_\ell\times W_\ell}$, which capture
local spatial variation along the two image axes. The monogenic
signal~\cite{felsberg2001} then defines the local amplitude and phase as
\begin{equation}
A = \sqrt{b^2 + R_1^2 + R_2^2}, \
\phi = \operatorname{atan2}\left(
\sqrt{R_1^2+R_2^2},b\right).
\label{eq:monogenic}
\end{equation}
All operations are pointwise over $(x,y)$, so
$A,\phi\in\mathbb{R}^{H_\ell\times W_\ell}$. The amplitude measures local
contrast, while the phase encodes the spatial position of structure.

\subsection{Phase-reliability adaptive gain}

Since $\phi$ in \eqref{eq:monogenic} is meaningful only when $A$ is
appreciable, we define a reliability weight
\begin{equation}
\rho = \frac{A^2}{A^2 + \sigma^2},
\label{eq:rho}
\end{equation}
a Wiener-type shrinkage with $\rho \to 1$ near strong edges and
$\rho \to 0$ in flat regions. Amplitudes well below $\sigma$ therefore
contribute little to the output. The threshold $\sigma$ is set to the
$75$th percentile of the amplitude distribution of the band itself, which
lets it adapt to the band content and removes the need for a hand-tuned
global constant.

The effective gain is
\begin{equation}
\alpha_{\text{eff}} = \alpha \, \frac{\rho}{1 + \nu\,(1-\rho)},
\label{eq:gain}
\end{equation}
which we use as a reliability-dependent shrinkage of the nominal gain. The
numerator suppresses amplification where the phase is unreliable, while the
denominator adds a penalty that grows with the operator order $\nu$ and
disappears as $\rho\to1$. The chosen form recovers the nominal gain $\alpha$
for fully reliable phase, suppresses the gain when the phase is unreliable,
and reduces to the plain weighting $\alpha\rho$ when $\nu=0$. Hence a larger
$\nu$, i.e. a stronger emphasis of high temporal frequencies, incurs a larger
penalty wherever the phase is unreliable.

The amplitude used in \eqref{eq:rho} is computed from the temporal mean of
each band, not from a single frame; a frame-by-frame mask would fluctuate with
the noise and introduce flicker.

\begin{figure*}[t]
\centering
\includegraphics[width=\textwidth]{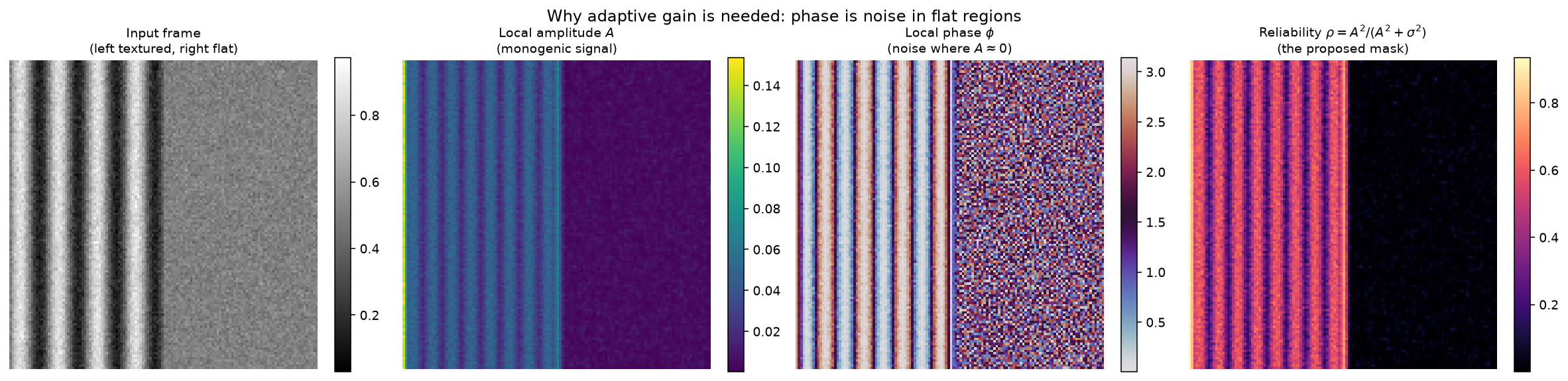}
\caption{Motivation for the adaptive gain. The input (left) is textured on the
left half and flat on the right. Local phase (third panel) is noise in the flat
region, where amplitude vanishes; the reliability mask \eqref{eq:rho} (right)
identifies this without supervision.}
\label{fig:mask}
\end{figure*}

\subsection{Algorithm}

\begin{algorithmic}[1]
\STATE \textbf{input:} frames $\{I_t\}_{t=1}^T$, $f_s$, band
$[f_{lo},f_{hi}]$, gain $\alpha$, order $\nu$, levels $L$
\FOR{$t = 1$ to $T$}
  \STATE $P_t \leftarrow \textsc{LaplacianPyramid}(I_t, L)$
\ENDFOR
\FOR{each pyramid band $\ell$}
  \IF{$\ell$ is the low-pass residual}
    \STATE leave unmodified \COMMENT{amplifying it would alter global scene content}
  \ELSE
    \STATE $S \leftarrow \textsc{Stack}(\{P_t[\ell]\}_t)$
    \STATE $\tilde{S} \leftarrow \textsc{FracBandPass}(S, f_s, f_{lo}, f_{hi}, \nu)$
    \STATE $A \leftarrow \textsc{Amplitude}(\textsc{Mean}_t(S))$
    \STATE $\alpha_{\text{eff}} \leftarrow$ \eqref{eq:gain} from $A$, $\nu$
    \STATE $P_t[\ell] \leftarrow P_t[\ell] + \alpha_{\text{eff}} \odot \tilde{S}_t
    \;\; \forall t$
  \ENDIF
\ENDFOR
\STATE \textbf{return} $\{\textsc{Collapse}(P_t)\}_t$
\end{algorithmic}

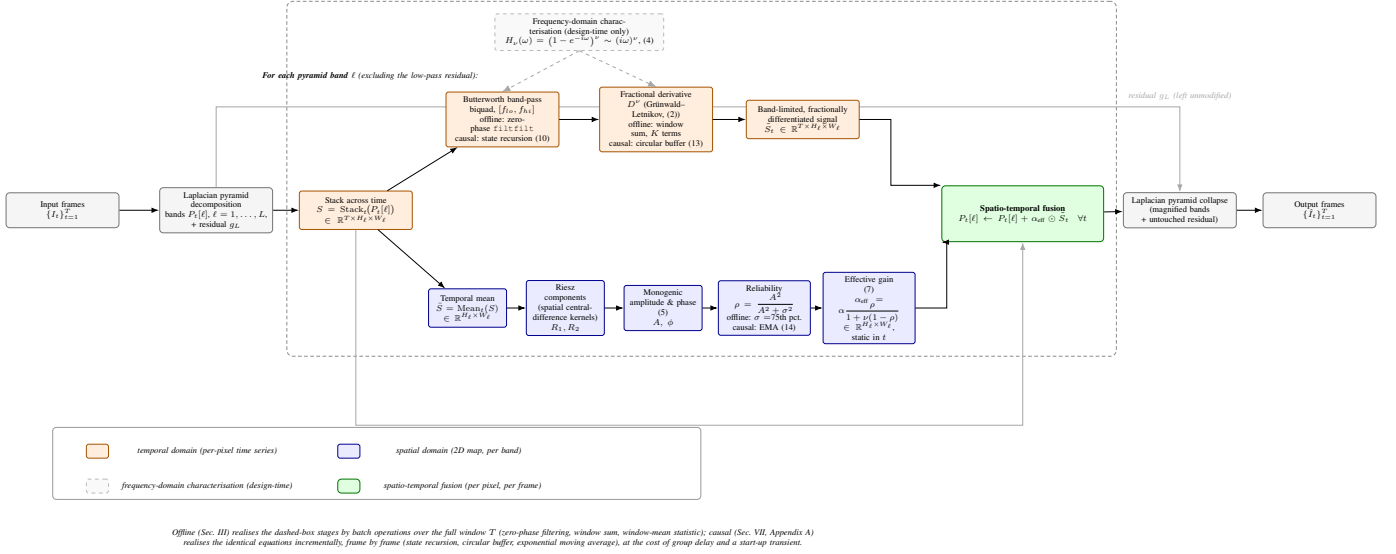
\begin{figure*}[t]
\centering
\resizebox{\textwidth}{!}{%
\begin{tikzpicture}[
  font=\scriptsize,
  >={Latex[length=2.2mm,width=1.6mm]},
  iostyle/.style={draw, thick, rounded corners=3pt, align=center,
    text width=3.0cm, minimum height=1.0cm, fill=gray!8, draw=black!55},
  temp/.style={draw, rounded corners=2pt, align=center,
    text width=3.0cm, minimum height=1.1cm, fill=orange!13, draw=orange!65!black},
  spat/.style={draw, rounded corners=2pt, align=center,
    text width=2.0cm, minimum height=1.1cm, fill=blue!8, draw=blue!55!black},
  freq/.style={draw, dashed, rounded corners=2pt, align=center,
    text width=4.6cm, minimum height=1.0cm, fill=gray!5, draw=gray!55},
  fuse/.style={draw, thick, rounded corners=3pt, align=center,
    text width=4.4cm, minimum height=1.6cm, fill=green!12, draw=green!45!black},
  lbl/.style={font=\scriptsize\itshape, align=center},
  grp/.style={draw=black!45, dash pattern=on 3pt off 2pt, rounded corners=5pt, inner sep=10pt}
]

\node[iostyle] (input) at (0,0) {Input frames\\$\{I_t\}_{t=1}^{T}$};
\node[iostyle] (pyr)   at (4.4,0) {Laplacian pyramid decomposition\\bands $P_t[\ell]$, $\ell=1,\dots,L$,\\+ residual $g_L$};
\node[iostyle] (rec)   at (32.0,0) {Laplacian pyramid collapse\\(magnified bands\\+ untouched residual)};
\node[iostyle] (outp)  at (36.0,0) {Output frames\\$\{\hat{I}_t\}_{t=1}^{T}$};

\draw[->,thick] (input) -- (pyr);
\draw[->,thick] (rec) -- (outp);

\draw[->,thick,gray!70] (pyr.north) |- ++(0,2.3) -| node[lbl,pos=0.5,above,yshift=2pt]{residual $g_L$ (left unmodified)} (rec.north);

\node[lbl, anchor=south west] at (5.6,3.6) {\textbf{For each pyramid band $\ell$} (excluding the low-pass residual):};

\node[temp] (stack) at (8.4,0) {Stack across time\\$S=\mathrm{Stack}_t\!\big(P_t[\ell]\big)$\\$\in\mathbb{R}^{T\times H_\ell\times W_\ell}$};

\draw[->,thick] (pyr) -- (stack);

\node[temp] (bp)   at (12.6,2.6) {Butterworth band-pass\\biquad, $[f_{lo},f_{hi}]$\\{\scriptsize offline: zero-phase \texttt{filtfilt}\\causal: state recursion \eqref{eq:biquad}}};
\node[temp] (gl)   at (17.0,2.6) {Fractional derivative\\$D^{\nu}$ (Gr\"unwald--Letnikov, \eqref{eq:gl})\\{\scriptsize offline: window sum, $K$ terms\\causal: circular buffer \eqref{eq:gl-inc}}};
\node[temp] (stil) at (21.2,2.6) {Band-limited, fractionally\\differentiated signal\\$\tilde{S}_t\in\mathbb{R}^{T\times H_\ell\times W_\ell}$};

\draw[->,thick] (stack) -- (bp);
\draw[->,thick] (bp) -- (gl);
\draw[->,thick] (gl) -- (stil);

\node[freq] (freqbox) at (14.8,5.1) {Frequency-domain characterisation (design-time only)\\$H_{\nu}(\omega)=\left(1-e^{-i\omega}\right)^{\nu}\sim(i\omega)^{\nu}$, \eqref{eq:response}};
\draw[->,dashed,gray!70] (freqbox.south) -- (bp.north);
\draw[->,dashed,gray!70] (freqbox.south) -- (gl.north);

\node[spat] (mean) at (11.6,-2.8) {Temporal mean\\$\bar{S}=\mathrm{Mean}_t(S)$\\$\in\mathbb{R}^{H_\ell\times W_\ell}$};
\node[spat] (riesz) at (14.4,-2.8) {Riesz\\components\\(spatial central-\\difference kernels)\\$R_1,R_2$};
\node[spat] (mono) at (17.2,-2.8) {Monogenic\\amplitude \& phase\\\eqref{eq:monogenic}\\$A,\ \phi$};
\node[spat, text width=2.4cm] (rho) at (20.1,-2.8) {Reliability\\$\rho=\dfrac{A^2}{A^2+\sigma^2}$\\{\scriptsize offline: $\sigma=$75th pct.\\causal: EMA \eqref{eq:ema}}};
\node[spat, text width=2.4cm] (gain) at (23.1,-2.8) {Effective gain\\\eqref{eq:gain}\\$\alpha_{\text{eff}}=\alpha\dfrac{\rho}{1+\nu(1-\rho)}$\\$\in\mathbb{R}^{H_\ell\times W_\ell}$, static in $t$};

\draw[->,thick] (stack) -- (mean);
\draw[->,thick] (mean) -- (riesz);
\draw[->,thick] (riesz) -- (mono);
\draw[->,thick] (mono) -- (rho);
\draw[->,thick] (rho) -- (gain);

\node[fuse] (fuse) at (27.5,-0.1) {\textbf{Spatio-temporal fusion}\\$P_t[\ell]\leftarrow P_t[\ell]+\alpha_{\text{eff}}\odot\tilde{S}_t\quad\forall t$};

\draw[->,thick] (stil.east) -- ++(0.9,0) |- (fuse.north west);
\draw[->,thick] (gain.east) -- ++(0.9,0) |- (fuse.south west);
\draw[->,thick,gray!70] (stack.south) -- ++(0,-5.6) -| node[lbl,pos=0.15,below,xshift=-1.1cm]{original band $P_t[\ell]$} (fuse.south);

\draw[->,thick] (fuse) -- (rec);

\begin{scope}[on background layer]
\node[grp, fit=(freqbox)(bp)(gl)(stil)(mean)(riesz)(mono)(rho)(gain)(fuse)(stack)] {};
\end{scope}

\node[draw=black!40, rounded corners=3pt, fill=white, inner sep=8pt,
      anchor=north west, fit={(0,-6.5) (18,-8.0)}] (legendbg) {};

\node[temp, text width=0.4cm, minimum height=0.4cm] (lg1) at (1.0,-6.9) {};
\node[lbl, anchor=west, text width=5.4cm] at (lg1.east) {temporal domain (per-pixel time series)};
\node[spat, text width=0.4cm, minimum height=0.4cm] (lg2) at (8.2,-6.9) {};
\node[lbl, anchor=west, text width=5.4cm] at (lg2.east) {spatial domain (2D map, per band)};
\node[freq, text width=0.4cm, minimum height=0.4cm] (lg3) at (1.0,-7.9) {};
\node[lbl, anchor=west, text width=5.4cm] at (lg3.east) {frequency-domain characterisation (design-time)};
\node[fuse, text width=0.4cm, minimum height=0.4cm] (lg4) at (8.2,-7.9) {};
\node[lbl, anchor=west, text width=5.6cm] at (lg4.east) {spatio-temporal fusion (per pixel, per frame)};

\node[lbl, anchor=north west, text width=24cm] at (0.2,-9.0)
  {\textit{Offline} (Sec.~III) realises the dashed-box stages by batch
  operations over the full window $T$ (zero-phase filtering, window sum,
  window-mean statistic); \textit{causal} (Sec.~VII, Appendix~\ref{app:impl}) realises the
  identical equations incrementally, frame by frame (state recursion,
  circular buffer, exponential moving average), at the cost of group delay
  and a start-up transient.};

\end{tikzpicture}%
}
\caption{End-to-end data flow of FrAM for one pyramid band $\ell$. The loop
over bands runs independently for every scale, and the low-pass residual
bypasses it completely. After the temporal stack $S$ the pipeline splits into
a temporal/frequency branch that produces the fractionally differentiated
signal $\tilde{S}_t$, and a spatial branch computed once from the band's
temporal mean that produces the static gain map $\alpha_{\text{eff}}$; the two
are fused with the original unfiltered band. Stages with both offline and
causal realisations are annotated with both modes
(Sec.~VII, Appendix~\ref{app:impl}).}
\label{fig:pipeline}
\end{figure*}

Here, $\ell$ indexes the spatial bands of the Laplacian pyramid. The low-pass
residual is the coarsest, spatially smoothed component, containing large-scale
scene structure and slowly varying intensity. It is left unchanged because
magnifying it could alter overall scene appearance or illumination rather than
the targeted temporal variations. Fig.~\ref{fig:pipeline} summarises the
complete data flow, distinguishing spatial, temporal, frequency-domain, and
fusion stages.

\section{Experimental Setup}

\subsection{Why a synthetic sequence}
Quantifying amplification requires knowing the true displacement, information
that no real video supplies at sub-pixel precision. We therefore use a
controlled sequence that also lets the two questions of interest be separated:
how much genuine motion is amplified, and how much noise is injected where
there is no signal to amplify.

The sequence has $96$ frames of $128\times128$ at $30$\,fps. Its left half
contains a horizontal sinusoid of period $16$\,px that oscillates sideways
with amplitude $0.2$\,px at $2$\,Hz; the right half is \emph{flat}. Additive
Gaussian noise of standard deviation $\sigma_{\text{in}}$ is added
throughout. The displacement is well below one pixel and cannot be seen by
inspection. The temporal pass-band used in the experiments is
$[1.5,2.5]$\,Hz, so the $2$\,Hz ground-truth oscillation lies at its centre.
Unless stated otherwise, all processing is performed on a single luminance
channel (grayscale); the real-time colour variant processes only the $Y$
channel of YCrCb.

\subsection{Metrics}
\textbf{Amplification factor}: motion amplification is quantified by comparing
the displacement estimated from the processed video with the known
ground-truth displacement. For each frame, displacement is estimated with
aggregate 1-D optical flow relative to the mean frame. Under brightness
constancy and for small displacements, $I_t + d I_x \approx 0$; the
least-squares solution over all pixels is
\[
d = -\frac{\sum I_t I_x}{\sum I_x^2}.
\]
After removing the temporal mean from each series, a through-origin
least-squares fit
\[
d_{\mathrm{est}} = a\,d_{\mathrm{GT}},
\qquad
a = \frac{\sum d_{\mathrm{GT}}\,d_{\mathrm{est}}}
         {\sum d_{\mathrm{GT}}^2},
\]
defines the \textbf{amplification factor} as the slope $a$. Thus $a=1$ means
faithful motion estimation, $a>1$ means amplification, and $a<1$ means
attenuation. Applied to the unprocessed input, the estimator gives $a=1.01$,
indicating a negligible intrinsic bias.

\textbf{Regional temporal noise}: to separate amplified sensor noise from
amplified motion, temporal intensity fluctuation is measured independently in
the flat and textured halves. Let
$\Delta I_t(\mathbf{x}) = I_{t+1}(\mathbf{x}) - I_t(\mathbf{x})$ be the
inter-frame difference. Over a spatial region $\mathcal{R}$ the noise score
is the pooled standard deviation
\[
\sigma_{\mathcal{R}}
  = \mathrm{std}\bigl\{\Delta I_t(\mathbf{x})
      :\, t=1,\ldots,T{-}1,\; \mathbf{x}\in\mathcal{R}\bigr\}.
\]
The flat region uses $\mathcal{R}_{\mathrm{flat}}=\{x\geq 70\}$ and the
textured region $\mathcal{R}_{\mathrm{tex}}=\{x<58\}$, leaving a margin
around the texture--flat boundary at $x=64$. Consecutive-frame differencing
high-pass filters the time series, so $\sigma_{\mathcal{R}}$ emphasises
temporal mottling rather than static spatial texture; in the flat half it is
therefore a direct readout of injected noise.

\textbf{Band SNR}: spectral fidelity of the recovered motion is measured on
the same displacement series $d_{\mathrm{est}}(t)$ used for the amplification
factor. After mean removal we form the one-sided periodogram
$P(f)=|\widehat{d}_{\mathrm{est}}(f)|^2$ via the real DFT and define
\[
\begin{aligned}
\mathrm{SNR}_{\mathrm{band}}
  &= 10\log_{10}\!\frac{P_{\mathrm{in}}}{P_{\mathrm{out}}},\\
P_{\mathrm{in}}
  &= \sum_{|f-f_s|<\Delta f} P(f),\\
P_{\mathrm{out}}
  &= \sum_{\substack{f>0\\ |f-f_s|\geq\Delta f}} P(f),
\end{aligned}
\]
with signal frequency $f_s=2$\,Hz and half-width $\Delta f=0.5$\,Hz, matching
the synthetic oscillation. Spatial intensity variance is deliberately
\emph{not} used: under a near-rigid translation it is almost constant in
time and carries no oscillatory signature. A higher
$\mathrm{SNR}_{\mathrm{band}}$ indicates that the magnified motion
concentrates energy at $f_s$ rather than spreading it into broadband temporal
noise.

\subsection{Protocol}
Methods are compared with $\alpha$ \textbf{calibrated toward equal effective
amplification}. Because the pipeline is linear in $\alpha$ under a spatially
uniform gain, each configuration's $\alpha$ is scaled by the ratio of the
baseline amplification to a probe run. With the reliability mask the realised
factor can remain slightly below the baseline, because flat pixels receive
little gain. This is the only fair comparison: any method would look less
noisy if it simply amplified less.

\section{Results}

\begin{table}[t]
\caption{Comparison with $\alpha$ calibrated toward equal effective
amplification (EVM $18.44\times$), $\sigma_{\text{in}} = 0.04$.
FrAM falls slightly short of the EVM factor because the reliability mask
withholds gain in the flat half; lower noise is better. Bold marks the better
value of each metric column among the processed methods (noise, SNR); the
amplification serves as the calibration target, and the unprocessed input is
listed for reference only.}
\label{tab:main}
\centering
\begin{tabular}{lccccc}
\toprule
Method & $\alpha$ & Amplif. & Noise & Noise & SNR \\
       &          &         & (flat) & (textured) & (dB) \\
\midrule
Input (unprocessed)   & --  & 1.01$\times$  & 0.0565 & 0.0569 & 8.86 \\
EVM (Butterworth)     & 20  & 18.44$\times$ & 0.1137 & 0.1589 & \textbf{11.99} \\
\midrule
FrAM ($\nu = 0.0$)    & 42  & 17.03$\times$ & \textbf{0.0574} & \textbf{0.1525} & 11.89 \\
FrAM ($\nu = 0.3$)    & 64  & 16.09$\times$ & \textbf{0.0574} & 0.1589 & 11.91 \\
FrAM ($\nu = 0.5$)    & 90  & 14.83$\times$ & 0.0575 & 0.1733 & 11.74 \\
FrAM ($\nu = 0.7$)    & 131 & 12.77$\times$ & 0.0576 & 0.1987 & 11.41 \\
\bottomrule
\end{tabular}
\end{table}

\subsection{Main result}
Table~\ref{tab:main} reports the main comparison. With $\alpha$ calibrated
toward the Eulerian factor, FrAM holds flat-region temporal noise at the input
level ($0.0574$, versus $0.0565$ for the input), while the Eulerian baseline
doubles it to $0.1137$; the output therefore remains clean in regions where
the baseline injects spurious motion. The residual gap in measured
amplification ($17.0\times$ at $\nu=0$ versus $18.4\times$) is the expected
cost of withholding gain where there is no structure to amplify.

The textured region behaves differently depending on $\nu$. At
$\nu \leq 0.3$ FrAM matches the baseline ($0.1589$ for both at $\nu=0.3$);
raising $\nu$ increases textured-region noise, as \eqref{eq:response}
predicts, because higher order emphasises the high frequencies where noise
dominates. Together with the loss of amplification at high $\nu$, this points
to $\nu \in [0, 0.3]$ as the useful operating range.

\subsection{Ablation}

\begin{table}[t]
\caption{Adaptive-gain ablation, $\alpha = 20$ fixed. Within each order, bold
marks the gain with the lower noise in each noise column. At fixed $\alpha$
the adaptive gain always lowers noise and always lowers amplification;
matching amplification requires recalibrating $\alpha$
(Table~\ref{tab:main}).}
\label{tab:ablation}
\centering
\begin{tabular}{llccc}
\toprule
Order & Gain & Amplif. & Noise (flat) & Noise (textured) \\
\midrule
$\nu = 0.0$ & constant & 18.44$\times$ & 0.1137 & 0.1589 \\
$\nu = 0.0$ & adaptive & 8.75$\times$  & \textbf{0.0569} & \textbf{0.0925} \\
$\nu = 0.5$ & constant & 10.05$\times$ & 0.0935 & 0.1188 \\
$\nu = 0.5$ & adaptive & 4.11$\times$  & \textbf{0.0567} & \textbf{0.0723} \\
$\nu = 0.7$ & constant & 6.70$\times$  & 0.0876 & 0.1072 \\
$\nu = 0.7$ & adaptive & 2.81$\times$  & \textbf{0.0566} & \textbf{0.0679} \\
\bottomrule
\end{tabular}
\end{table}

Table~\ref{tab:ablation} separates the contribution of the two components.
With a constant gain, increasing $\nu$ does reduce noise, but at a severe cost
in amplification ($18.44\times \to 6.70\times$). The adaptive gain, by
contrast, brings flat-region noise down to the input level at every order.
The noise reduction is therefore attributable to the adaptive gain, not to
the fractional order, whose role is to control the amplification--noise
trade-off in textured regions.

\subsection{Robustness}

\begin{table}[t]
\caption{Robustness to input noise, at matched amplification, $\nu = 0.3$.
Bold marks the lower flat-region noise of the two methods on each row; the
amplification columns record the calibration target, not a quantity to
maximise.}
\label{tab:robust}
\centering
\begin{tabular}{ccccc}
\toprule
$\sigma_{\text{in}}$ & \multicolumn{2}{c}{EVM} & \multicolumn{2}{c}{FrAM} \\
\cmidrule(lr){2-3} \cmidrule(lr){4-5}
 & Amplif. & Noise (flat) & Amplif. & Noise (flat) \\
\midrule
0.01 & 19.15$\times$ & 0.0284 & 16.81$\times$ & \textbf{0.0142} \\
0.02 & 18.98$\times$ & 0.0569 & 16.64$\times$ & \textbf{0.0284} \\
0.04 & 18.44$\times$ & 0.1137 & 16.09$\times$ & \textbf{0.0574} \\
0.08 & 16.64$\times$ & 0.2275 & 14.36$\times$ & \textbf{0.1207} \\
\bottomrule
\end{tabular}
\end{table}

Table~\ref{tab:robust} shows that the behaviour is stable across an eightfold
range of input noise: FrAM roughly halves flat-region noise at every level
while retaining approximately $87\%$ of the baseline amplification.

\begin{figure*}[t]
\centering
\includegraphics[width=\textwidth]{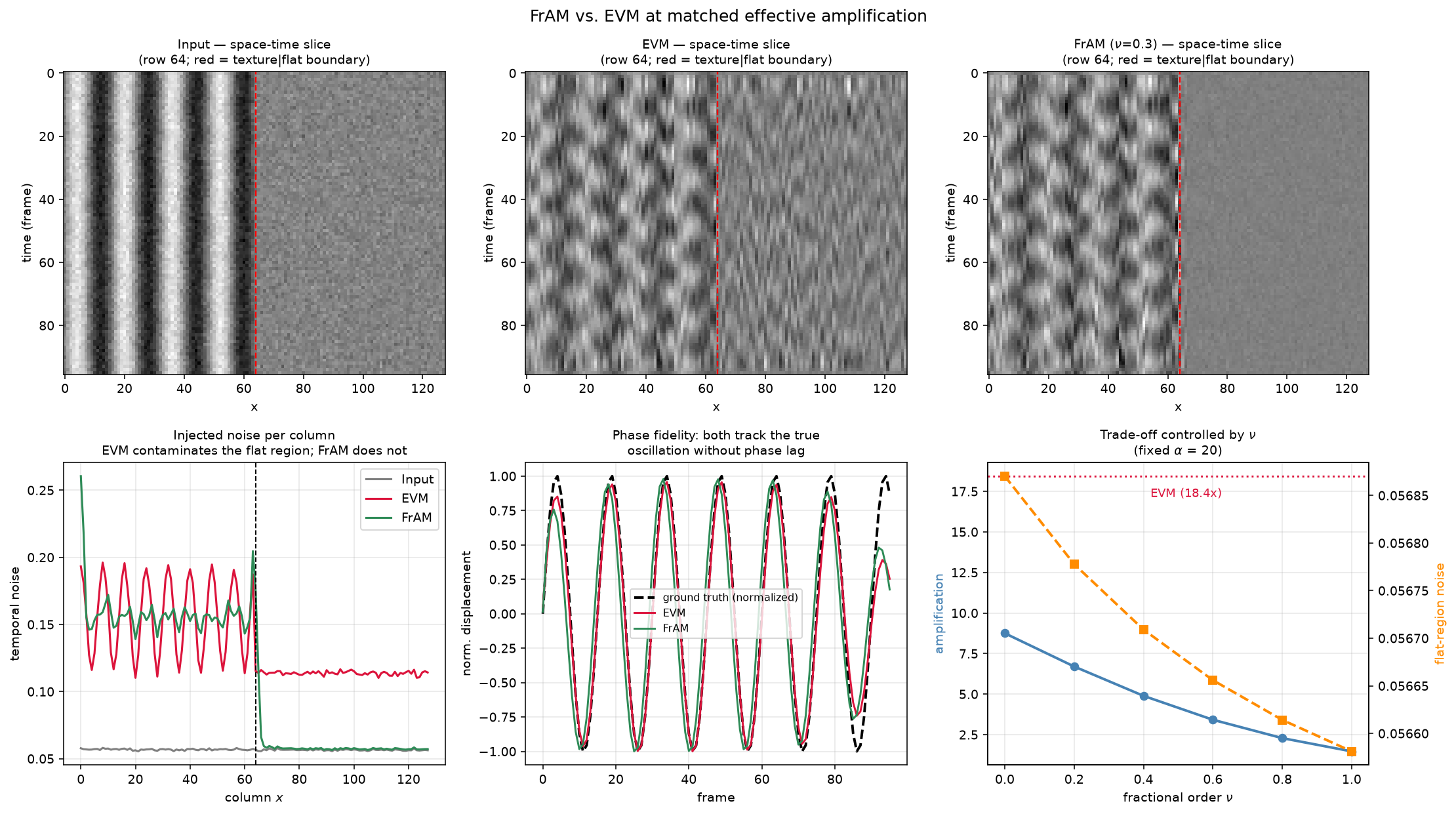}
\caption{Comparison at equal effective amplification. Top: spatio-temporal
slices; right of the dashed line the region is flat, and the baseline has filled
it with spurious texture that FrAM leaves untouched. Bottom left: injected noise
per column, dropping sharply at the boundary for FrAM. Bottom centre: both
methods track the true oscillation without phase lag. Bottom right: the
trade-off controlled by $\nu$.}
\label{fig:results}
\end{figure*}

Fig.~\ref{fig:results} presents the same result visually. The spatio-temporal
slices show the flat region of the baseline output populated with spurious
temporal texture, absent from FrAM. The displacement traces confirm that
neither method introduces phase lag, as expected from zero-phase filtering.

\section{Validation on Real Video}

The synthetic experiment quantifies behaviour under controlled conditions but
does not establish that the method is useful on real material. We therefore
evaluate on the four sequences distributed with the reference implementation of
\cite{wu2012}: \texttt{face} and \texttt{face2} (facial pulse), \texttt{baby}
(infant respiration) and \texttt{subway} (structural vibration). Each clip was
processed as a single grayscale channel at half resolution (or $0.35\times$ for
all but \texttt{face}), using the first $180$ frames.

\subsection{Evaluation without ground truth}

Real video provides no ground-truth sub-pixel displacement, so the
amplification factor of Section~IV cannot be computed. We substitute three
criteria that remain objective:

\textbf{(1) Physiological selectivity.} In the facial sequences the pulse lies
in a known band ($0.7$--$2.0$\,Hz, the band used for
\texttt{face}/\texttt{face2}; \texttt{baby} uses $0.5$--$2.0$\,Hz and
\texttt{subway} $1.0$--$4.0$\,Hz). We report the band energy ratio (BER),
the fraction of temporal energy falling inside the target band, and the
dominant frequency in beats per minute.

\textbf{(2) Noise in signal-free regions.} Each scene is segmented
\emph{without supervision} into an active and a static region by thresholding
the per-pixel temporal standard deviation at the 80th and 30th percentiles.
The masks are computed on the \emph{original} video, so they are
identical for every method compared. Any temporal noise injected into the
static region is pure artefact.

\textbf{(3) Spatial selectivity.} The ratio of enhancement in the active region
to enhancement in the static region. A value of unity indicates
indiscriminate amplification.

Methods are compared at \textbf{equal enhancement in the active region}:
FrAM's $\alpha$ is scaled until its enhancement ratio in the active region
matches the EVM one, as in the synthetic protocol.

\begin{table}[t]
\caption{Real video, at equal enhancement in the active region. Within each
video, bold marks the better of EVM and FrAM on each metric (higher BER, lower
background noise, higher spatial selectivity); EVM retains slightly higher BER
on three of the four clips. The unprocessed input is shown once as a
reference, and the pulse rate is a consistency check rather than a quantity to
maximise.}
\label{tab:real}
\centering
\begin{tabular}{llcccc}
\toprule
Video & Method & BER & Noise & Spatial & Rate \\
      &        &     & (bg)  & select. & (bpm) \\
\midrule
\texttt{face} & input & 0.136 & 0.00233 & 1.00 & 50 \\
\texttt{face} & EVM   & 0.825 & 0.00909 & 1.40 & 50 \\
\texttt{face} & FrAM  & \textbf{0.844} & \textbf{0.00302} & \textbf{4.53} & 50 \\
\midrule
\texttt{face2} & EVM  & \textbf{0.332} & 0.00702 & 1.14 & 50 \\
\texttt{face2} & FrAM & 0.244 & \textbf{0.00662} & \textbf{1.31} & 70 \\
\midrule
\texttt{baby} & EVM   & \textbf{0.395} & 0.00468 & 1.65 & -- \\
\texttt{baby} & FrAM  & 0.362 & \textbf{0.00246} & \textbf{3.49} & -- \\
\midrule
\texttt{subway} & EVM  & \textbf{0.742} & 0.02026 & 1.16 & -- \\
\texttt{subway} & FrAM & 0.598 & \textbf{0.00780} & \textbf{3.19} & -- \\
\bottomrule
\end{tabular}
\end{table}

\subsection{Results}

Table~\ref{tab:real} shows that FrAM reduces background noise on all four
sequences, by a margin that varies considerably. On \texttt{face} the reduction
is $3.0\times$ ($0.00302$ versus $0.00909$) while in-band energy is preserved
($\text{BER} = 0.844$ versus $0.825$) and the same pulse rate is recovered;
spatial selectivity rises from $1.40$ to $4.53$. On \texttt{subway} background
noise falls by $2.6\times$ and selectivity from $1.16$ to $3.19$. On
\texttt{baby} the reduction is $1.9\times$. On \texttt{face2} the advantage is
marginal ($1.06\times$).

The consistent pattern across sequences is spatial selectivity: FrAM raises it
in every case, most sharply on \texttt{face} and \texttt{subway}. This is the
behaviour the reliability weighting is designed to produce, and it is the
quantity least sensitive to how the enhancement is calibrated.

\begin{table}[t]
\caption{Activity contrast of the input against the observed advantage. Bold
marks the lower background noise of the two methods on each row; the gain
column is the ratio of the two noise columns. The ordering is not monotonic:
contrast does not predict the margin.}
\label{tab:contrast}
\centering
\begin{tabular}{lcccc}
\toprule
Video & Activity & Noise (bg) & Noise (bg) & Gain \\
      & contrast & EVM        & FrAM       &      \\
\midrule
\texttt{subway} & 6.04 & 0.02026 & \textbf{0.00780} & $2.60\times$ \\
\texttt{face2}  & 3.77 & 0.00702 & \textbf{0.00662} & $1.06\times$ \\
\texttt{face}   & 1.98 & 0.00909 & \textbf{0.00302} & $3.01\times$ \\
\texttt{baby}   & 1.07 & 0.00468 & \textbf{0.00246} & $1.90\times$ \\
\bottomrule
\end{tabular}
\end{table}

Define the \emph{activity contrast} of a sequence as the ratio of temporal
noise in the active region to that in the static region of the \emph{original}
video. Table~\ref{tab:contrast} shows that this quantity does \emph{not} order
the outcomes: the largest gain occurs at a contrast of $1.98$ and the smallest
at $3.77$, and the linear correlation over the four sequences is negligible
($r = 0.03$). Four sequences cannot support a predictor of the margin, and we
offer none.

The advantage is smallest on \texttt{face2}, whose input carries the least
in-band energy of the four ($\text{BER} = 0.022$, roughly six times below the
other three). Where the band-pass isolates little genuine signal, both
methods amplify predominantly noise and the reliability mask has
correspondingly little to distinguish. We regard this as a plausible
mechanism, not a validated criterion: settling an applicability rule would
require far more than four clips.

\begin{figure*}[t]
\centering
\includegraphics[width=\textwidth]{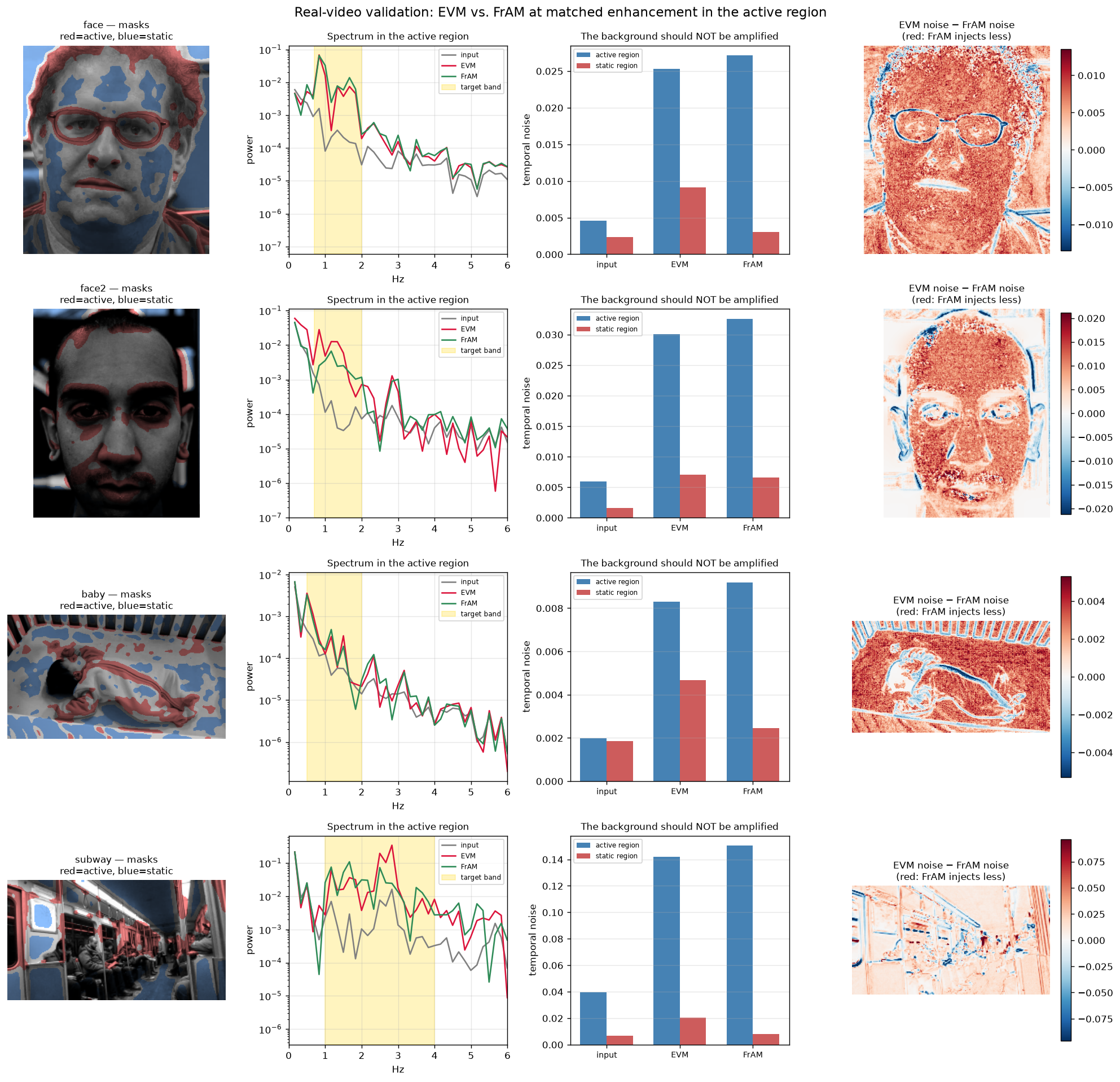}
\caption{Real-video validation. Columns: unsupervised region masks; temporal
spectrum of the active region with the target band shaded; temporal noise per
region; and the difference map of injected noise (red where FrAM injects less).
Red dominates on all four sequences, most clearly for \texttt{face} and
\texttt{subway}.}
\label{fig:real}
\end{figure*}

\section{Real-Time Formulation}

\subsection{Diagnosis of the cost}

The pipeline of Section~III, applied to a sliding window, costs over
$2$\,s per output frame at $640\times480$ with $T=32$ ($2430$\,ms in
Table~\ref{tab:speed}). Profiling attributes the time almost entirely to the
temporal filter:

\begin{center}
\begin{tabular}{lr}
\toprule
Stage & Cost \\
\midrule
Laplacian pyramids ($T=32$ frames) & $\sim$90\,ms \\
\textbf{Temporal filter} & $\mathbf{\sim}$1.95\,s \\
Pyramid collapse ($T=32$ frames) & $\sim$50\,ms \\
\bottomrule
\end{tabular}
\end{center}

Three sources of inefficiency compound here. First, the window is reprocessed
in full at every step and all but the newest frame is discarded. Second,
\texttt{filtfilt} traverses the sequence twice to obtain zero phase. Third,
the convolution \eqref{eq:gl} is evaluated as $K$ shifts of the whole
$T \times H \times W$ volume.

\subsection{Causal reformulation}

None of the three is inherent to fractional calculus; all arise from
processing a window instead of a stream. The same operator admits a causal,
incremental form built on three substitutions, detailed in
Appendix~\ref{app:impl}:

\begin{enumerate}
\item The band-pass of Section~III is evaluated as a second-order state
recursion \eqref{eq:biquad}, keeping two past inputs and outputs per pixel.
This is the identical filter, not an approximation: against \texttt{lfilter}
the maximum discrepancy is $6\times10^{-16}$, i.e.\ machine precision. Cost
per frame drops from $O(T)$ to $O(1)$.
\item The truncated Gr\"unwald--Letnikov sum \eqref{eq:gl} is evaluated over a
circular buffer of $K+1$ band images, as one weighted combination rather than
$K$ shifts of a $T \times H \times W$ volume.
\item The reliability statistic of \eqref{eq:rho} is updated by an
exponential moving average instead of being recomputed over the window mean.
\end{enumerate}

Causality carries two costs: the filter has non-zero phase, so it introduces
group delay where the offline mode has none, and roughly $40$ frames of
transient pass before the output is valid.

\begin{table}[t]
\caption{Per-frame cost: offline versus causal streaming. Streaming is
cheaper at every resolution tested; bold highlights the $640\times480$ row,
the reference configuration quoted elsewhere in this paper.}
\label{tab:speed}
\centering
\begin{tabular}{lrrrr}
\toprule
Resolution & Offline & Streaming & Speed-up & fps \\
\midrule
$320\times240$   & 382\,ms   & 6.3\,ms   & $60\times$  & 158 \\
$640\times480$   & 2430\,ms  & 14.6\,ms  & $\mathbf{166\times}$ & \textbf{69} \\
$1280\times720$  & 7772\,ms  & 62.6\,ms  & $124\times$ & 16 \\
$1920\times1080$ & 17937\,ms & 108.8\,ms & $165\times$ & 9 \\
\bottomrule
\end{tabular}
\end{table}

\subsection{Results}

Table~\ref{tab:speed} reports the outcome: a $166\times$ speed-up at
$640\times480$ ($69$\,fps), crossing the $30$\,fps threshold at QVGA and VGA. A
colour variant operating on the luminance channel of YCrCb, leaving chrominance
untouched, costs $18.5$\,ms ($54$\,fps at $640\times480$): the magnification
work is confined to one channel, so its cost remains close to grayscale.
The timings in Table~\ref{tab:speed} come from a single run and are
representative rather than exact: repeated executions on the same machine give
values that can differ by tens of percent. The dependable outcome is the
order-of-magnitude gap between the two modes, not any individual ratio.

\begin{table}[t]
\caption{Streaming, at equal effective amplification. Bold marks the lowest
noise of each column among the processed methods; the unprocessed input is
listed for reference only.}
\label{tab:rt}
\centering
\begin{tabular}{lrrrr}
\toprule
Method & $\alpha$ & Amplif. & Noise (flat) & Noise (text.) \\
\midrule
Input            & --  & 1.01$\times$  & 0.0565 & 0.0569 \\
EVM (streaming)  & 20  & 20.85$\times$ & 0.1605 & 0.2002 \\
FrAM ($\nu=0.0$)  & 33  & 20.08$\times$ & \textbf{0.1091} & \textbf{0.1827} \\
FrAM ($\nu=0.15$) & 39  & 19.78$\times$ & 0.1149 & 0.1887 \\
FrAM ($\nu=0.3$)  & 47  & 19.36$\times$ & 0.1250 & 0.2026 \\
FrAM ($\nu=0.5$)  & 64  & 18.47$\times$ & 0.1480 & 0.2375 \\
\bottomrule
\end{tabular}
\end{table}

Table~\ref{tab:rt} confirms that the advantage survives the reformulation,
attenuated: flat-region noise falls from $0.1605$ to $0.1091$, a factor of
$1.47\times$ against the $2\times$ of the offline mode. The cause is that the
causal mask, lacking the average over the full window, is noisier than its
offline counterpart.

\begin{table}[t]
\caption{Truncating the fractional memory ($640\times480$). Amplification and
flat-region noise vary little for $K \geq 8$, while cost grows with $K$; bold
highlights the truncation $K = 8$ used by the interactive viewer.}
\label{tab:memory}
\centering
\begin{tabular}{rrrr}
\toprule
$K$ & Amplif. & Noise (flat) & ms/frame \\
\midrule
4  & 8.17$\times$ & 0.0779 & 7.8 \\
\textbf{8} & \textbf{8.94}$\times$ & \textbf{0.0783} & \textbf{11.2} \\
16 & 8.78$\times$ & 0.0781 & 12.1 \\
24 & 8.93$\times$ & 0.0782 & 15.0 \\
31 & 8.82$\times$ & 0.0781 & 17.9 \\
\bottomrule
\end{tabular}
\end{table}

Table~\ref{tab:memory} shows how much fractional memory is needed.
Truncating to $K=8$ yields the same amplification as $K=31$ at roughly
$60\%$ of its cost in this run. The long-memory property of the operator
matters for its spectral character (the $|\omega|^{\nu}$ response of
\eqref{eq:response}), but the far tail contributes little to the measured
result, so the memory noted in Sec.~III-A turns out to be largely unnecessary
in practice.

\begin{figure*}[t]
\centering
\includegraphics[width=\textwidth]{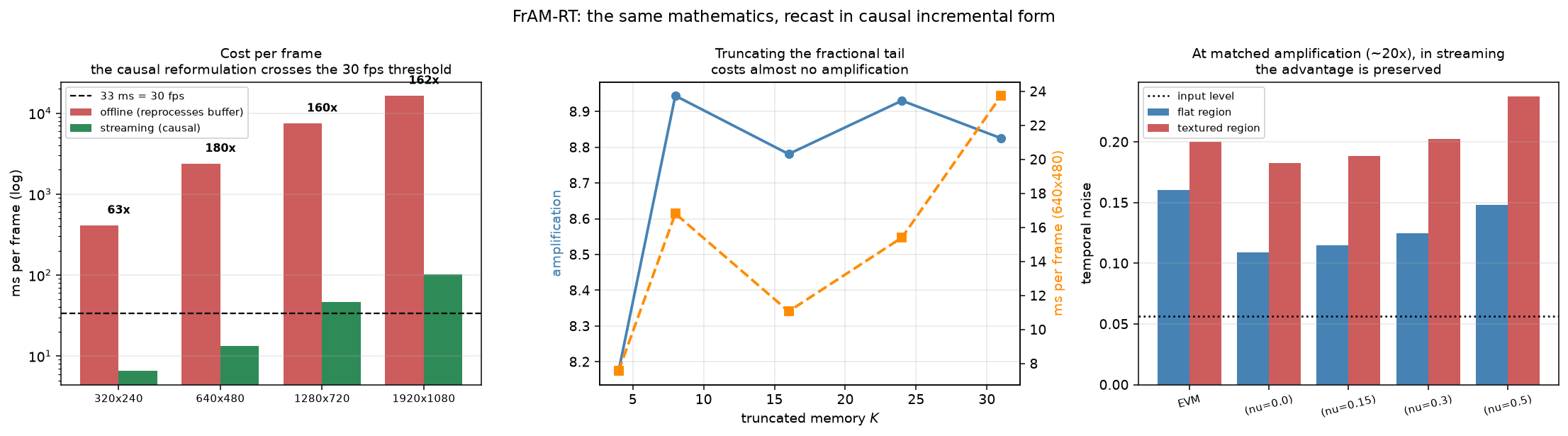}
\caption{Causal reformulation. Left: per-frame cost, log scale, with the
$30$\,fps threshold marked. Centre: truncating the fractional memory costs
little amplification while cost grows with $K$. Right: at equal amplification
in streaming, the advantage over the baseline is retained.}
\label{fig:rt}
\end{figure*}

\section{Limitations}

\textbf{Attribution.} As Table~\ref{tab:ablation} establishes, the noise
reduction follows from the adaptive gain; the fractional operator contributes
the continuous control parameter, not the improvement. A reader interested
only in the noise result could adopt \eqref{eq:gain} with $\nu = 0$.

\textbf{Variable margin, unexplained.} The background-noise reduction ranges
from $1.06\times$ to $3.01\times$ across four sequences, and
Table~\ref{tab:contrast} shows that the activity contrast of the input does not
predict where in that range a given sequence falls. We can offer a mechanism for
the weakest case but not a criterion, and four clips are too few to establish
one. Which scenes benefit most therefore remains open.

\textbf{Ambiguity on \texttt{face2}.} The gain is marginal and the dominant
frequency shifts from $50$ to $70$\,bpm. Both lie within the physiological
range, but the discrepancy should be resolved against a reference measurement
such as contact photoplethysmography.

\textbf{Two modes with different properties.} The offline mode uses a
zero-phase filter and yields the $2\times$ noise reduction; the causal mode
yields $1.47\times$. The figures reported for FrAM in this paper are those of
the offline mode and should not be quoted as streaming numbers.

\textbf{First-order Riesz approximation.} The kernels $[-0.5,0,0.5]$ are a
first-order approximation; a full steerable-filter implementation would estimate
phase more accurately.

\textbf{Global order.} A single $\nu$ is used for all bands. Since noise
characteristics differ across scales, a per-band order is a natural extension.

\section{Conclusion}

We presented FrAM, a motion-magnification method that combines a
fractional-order Gr\"unwald--Letnikov temporal filter with a gain weighted by
the reliability of the local phase. On a synthetic sequence with known
sub-pixel ground truth, the method approaches the amplification of classical
Eulerian magnification while holding temporal noise in textureless regions at
the input level, where the Eulerian baseline doubles it, and this holds
across an eightfold range of input noise. On real sequences the effect is
confirmed: background noise falls by $3.0\times$ on the primary facial-pulse
clip and $2.6\times$ on the structural-vibration clip, with in-band energy
and pulse rate preserved, and spatial selectivity improves on every sequence.
The noise reduction, however, ranges from $1.06\times$ to $3.01\times$ across
the four clips by a mechanism we could not identify. An ablation attributes
the improvement to the adaptive gain and characterises the fractional order
as a control parameter for the amplification--noise trade-off. The
reliability weighting is independent of the rest of the pipeline and could be
applied to existing phase-based magnification methods directly.

We further showed that the computational cost of the formulation was an
artefact of implementation rather than of the fractional operator. Recasting
the band-pass as a state recursion \eqref{eq:biquad} and the
Gr\"unwald--Letnikov sum over a circular buffer \eqref{eq:gl-inc} reduces the
per-frame cost by $166\times$, reaching $69$\,fps in grayscale and $54$\,fps in
colour at $640\times480$, and retains a $1.47\times$ noise reduction over the
baseline. Truncating the fractional memory to $K=8$ costs no
measurable amplification, which suggests the long-memory
property matters for the spectral character of the operator rather than for the
magnitude of the effect.

\section*{Reproducibility}

Source code, the interactive real-time viewer and result videos are available
at \expandafter\url\expandafter{\repourl}.

The implementation depends only on NumPy, SciPy, OpenCV and Matplotlib; no GPU,
deep-learning framework or specialised pyramid library is required. Every table
and figure in this paper is produced by one of three scripts, with fixed
parameters and fixed random seeds where randomness is involved:
\texttt{experiments.py} for the synthetic evaluation of
Tables~\ref{tab:main}--\ref{tab:robust}, \texttt{real\_video.py} for the
real-video validation of Tables~\ref{tab:real}--\ref{tab:contrast}, and
\texttt{experiments\_rt.py} for the streaming results of
Tables~\ref{tab:speed}--\ref{tab:memory}. The two magnification engines are
separate modules: \texttt{fracmag.py} implements the offline, zero-phase
formulation of Section~III, and \texttt{fracmag\_rt.py} the causal streaming
formulation of Section~VII and Appendix~\ref{app:impl}. The real sequences used
in Section~VI are those distributed with the reference implementation
of~\cite{wu2012}. All experiments process a single luminance or grayscale
channel; the real-time colour stream converts the input to YCrCb and magnifies
only the $Y$ channel.

\appendices

\section{Implementation of the Causal Formulation}
\label{app:impl}

This appendix collects the implementation details behind Section~VII. It is
separated from the body because it concerns realisation rather than method: the
operator being computed is the one defined in Section~III.

\subsection{Band-pass as a state recursion}

The band-pass filter of Section~III is a biquad. This follows from how
\texttt{scipy.signal.butter} realises a bandpass design: it builds an
order-$N$ lowpass \emph{prototype} and reaches the bandpass response through
the standard lowpass-to-bandpass substitution
\begin{equation}
s \;\longmapsto\; \frac{s^2+\omega_0^2}{sB},
\label{eq:lp2bp}
\end{equation}
where $\omega_0=\sqrt{\omega_{lo}\omega_{hi}}$ is the centre frequency and
$B=\omega_{hi}-\omega_{lo}$ the bandwidth. Substituting \eqref{eq:lp2bp} into
each simple pole of the prototype introduces an $s^2$ term in the
denominator, so a single pole becomes a conjugate \emph{pair} of poles: the
order doubles from $N$ to $2N$. For $N=1$, with prototype
$H_{lp}(s)=1/(s+1)$, this gives
\begin{equation}
H_{bp}(s) = H_{lp}\!\left(\frac{s^2+\omega_0^2}{sB}\right)
          = \frac{sB}{s^2+Bs+\omega_0^2},
\label{eq:bp-analog}
\end{equation}
an analog transfer function with two poles and one zero (at $s=0$). The
bilinear transform that discretises \eqref{eq:bp-analog} maps the point at
infinity to $z=-1$, contributing a second zero; the resulting discrete
transfer function therefore has two poles \emph{and} two zeros, i.e.\ three
coefficients $b_0,b_1,b_2$ and three coefficients $a_0,a_1,a_2$ ($a_0=1$
after normalisation); this is a biquad. This is why
\texttt{butter(1,~[lo,~hi],~'bandpass')}, despite the ``order 1'' argument,
returns coefficient arrays of length three rather than two: the argument
sets the order of the lowpass prototype, not of the realised bandpass
filter. Writing $b_i, a_i$ for these
coefficients, the output at frame $n$ is
\begin{equation}
y[n] = b_0 x[n] + b_1 x[n{-}1] + b_2 x[n{-}2] - a_1 y[n{-}1] - a_2 y[n{-}2],
\label{eq:biquad}
\end{equation}
evaluated elementwise over the band image, with the four state arrays
$x[n{-}1], x[n{-}2], y[n{-}1], y[n{-}2]$ retained between frames. Since
\eqref{eq:biquad} is the direct-form realisation of the same transfer function
implemented by \texttt{scipy.signal.lfilter}, the state recursion reproduces the
output of a single forward pass of that routine up to floating-point error; we
verified a maximum deviation of $6\times10^{-16}$ over a $60$-frame sequence.
The offline formulation uses \texttt{filtfilt}, which applies the same filter
forward and backward with carefully chosen initial conditions to achieve zero
phase; the causal streaming mode cannot replicate those initial conditions and
therefore introduces group delay.

\subsection{Fractional sum over a circular buffer}

This step does not change the Gr\"unwald--Letnikov derivative itself, only how
the $K+1$ frames it sums over are stored and addressed. Let $\mathbf{y}_n$
denote the band-pass output at frame $n$ and $w_k$ the weights of
\eqref{eq:recurrence}. As in \eqref{eq:gl}, the truncated sum needed at frame
$n$ is
\begin{equation}
D^{\nu} \mathbf{y}_n \approx \sum_{k=0}^{K} w_k \, \mathbf{y}_{n-k},
\label{eq:gl-window}
\end{equation}
i.e.\ it requires the current output together with its $K$ most recent
predecessors, $\mathbf{y}_n, \mathbf{y}_{n-1}, \dots, \mathbf{y}_{n-K}$, each
multiplied by its corresponding weight $w_0, w_1, \dots, w_K$. A na\"ive
sliding-window implementation keeps these $K+1$ frames as an explicit array
and shifts every entry down by one position after each frame, so that
$\mathbf{y}_{n-k}$ is always found at the same offset; this touches the whole
$T\times H\times W$ volume at every step, which is the cost the causal
formulation is meant to remove.

The circular buffer removes the shift by relaxing where each frame is
stored, and compensating with where it is read from. The $K+1$ frames are
kept in a fixed array of $K+1$ slots, $B[0],\dots,B[K]$, addressed through a
pointer $p$ that is updated in two steps whenever a new frame arrives:
$\mathbf{y}_n$ is written into the slot $p$ currently points to, and only
\emph{then} is $p$ advanced, wrapping back to $0$ once it would pass the last
slot: $p \leftarrow (p+1) \bmod (K{+}1)$, so that $p$ takes exactly the slot
indices $0,\dots,K$.

The same wrap-around is needed when reading a past frame, and gives the index
map below. Because writing happens \emph{before} advancing, the slot just
written, holding $\mathbf{y}_n$, sits at $(p-1)$ rather than at $p$; the frame
written the step before, $\mathbf{y}_{n-1}$, sits at $(p-2)$, and the $k$-th
most recent frame $\mathbf{y}_{n-k}$ at $(p-1-k)$. Taken as an ordinary
integer, $p-1-k$ can run negative (e.g.\ $p=0$, $k=2$ gives $-3$) or wrap past
$0$ repeatedly as $k$ grows, so it is folded back into the valid range
$\{0,\dots,K\}$ with the same $\bmod(K{+}1)$:
\begin{equation}
\mathbf{y}_{n-k} = B\big[(p - 1 - k) \bmod (K{+}1)\big],
\label{eq:gl-map}
\end{equation}
where $p$ is understood \emph{after} the write-then-advance step for frame
$n$. Concretely, with $K=3$ (four slots, $B[0..3]$): if the frame
$\mathbf{y}_n$ just written ends up leaving $p=0$ (it was written into
slot $3$, the last one, and $p$ then wrapped past it), the map
\eqref{eq:gl-map} places $\mathbf{y}_n$ at $B[3]$
($k{=}0$: $(0{-}1{-}0)\bmod4 = -1\bmod4 = 3$), $\mathbf{y}_{n-1}$ at $B[2]$
($k{=}1$: $-2\bmod4=2$), $\mathbf{y}_{n-2}$ at $B[1]$
($k{=}2$: $-3\bmod4=1$), and $\mathbf{y}_{n-3}$ at $B[0]$
($k{=}3$: $-4\bmod4=0$): the four stored frames are addressed backwards
from $p$, with no data ever copied between slots. Substituting \eqref{eq:gl-map} into
\eqref{eq:gl-window} gives
\begin{equation}
D^{\nu} \mathbf{y}_n \approx \sum_{k=0}^{K} w_k \, B\big[(p - 1 - k) \bmod (K{+}1)\big],
\label{eq:gl-inc}
\end{equation}
i.e.\ one weighted sum of $K+1$ stored images, with the index arithmetic
alone guaranteeing that the newest entry is always paired with $w_0$, the
second-newest with $w_1$, and so on, matching the ordering assumed in
\eqref{eq:gl-window}. Each new frame therefore costs one write into the slot
$p$ pointed to before it is advanced, followed by the weighted sum
\eqref{eq:gl-inc}, rather than $K$ shifts of a $T \times H \times W$ volume.
On the test machine the incremental cost of this weighted sum is a few
milliseconds per frame at $640\times480$ with $K=31$, far below the cost of
reprocessing the whole temporal window.

\subsection{Incremental reliability mask}

The offline mask of \eqref{eq:rho} takes its amplitude from the temporal mean of
each band, which is unavailable causally. It is replaced by an exponential
moving average,
\begin{equation}
\bar{A}_n = \tau \bar{A}_{n-1} + (1-\tau) A_n, \qquad \tau = 0.9,
\label{eq:ema}
\end{equation}
and the threshold $\sigma$ is refreshed from a percentile of $\bar{A}_n$ every
eighth frame and then itself smoothed; recomputing a percentile at every frame
would cost time and make the mask fluctuate visibly.

Because $\bar{A}_n$ is built from instantaneous bands rather than a temporal
mean, it carries more noise than its offline counterpart. To compensate, the
causal mask computes the quadrature pair with Sobel kernels, which factor as
$[1,2,1]^{\!\top}\ast[-1,0,1]$: up to a constant factor, the differencing part
coincides with the first-order Riesz kernels $[-0.5,0,0.5]$ of
\eqref{eq:monogenic}, while the transverse smoothing attenuates noise before
the amplitude is formed. This regularisation is a deliberate divergence from
the offline definition: the causal EMA of an instantaneous band needs the
extra smoothing to keep the mask stable. The cost of computing the mask is
sub-millisecond per frame at $640\times480$, comparable to that of the offline
Riesz-kernel variant.

\subsection{Numerical precision and colour}

All state and arithmetic are single precision. Measured on the recursion
\eqref{eq:biquad} at $640\times480$, \texttt{float32} costs $0.31$\,ms per frame
against $1.78$\,ms for \texttt{float64}, with no perceptible difference in the
output; the gain is bandwidth, not arithmetic.

The colour variant converts to YCrCb and applies the pipeline to the luminance
channel alone, leaving chrominance untouched apart from an optional saturation
factor. Since the magnification acts on geometry rather than colour, this
preserves the effect while keeping the cost close to the grayscale case
($18.5$\,ms for the colour stream against $14.6$\,ms for grayscale at
$640\times480$; run-to-run timing noise alone spans this difference).

\subsection{Parameter invalidation}

Amplification $\alpha$, the percentile, the adaptive flag and the method switch
enter only as scalars in the final combination, so they can change between
frames with no state rebuild. The band limits, the order $\nu$, the number of
pyramid levels, $K$, the colour mode and the frame size all determine filter
coefficients or buffer shapes, and changing any of them discards and rebuilds
the state. The viewer keys these on a tuple so that interactive slider
movements of the first group are free.

\end{document}